\documentclass[11pt, leqno]{article}

\usepackage{amsmath}
\usepackage{amsthm}
\usepackage{amssymb}
\usepackage[dvips]{graphicx,color}
\usepackage{bm}

\theoremstyle{plain}
\newtheorem{theorem}{Theorem}[section]
\newtheorem{lemma}[theorem]{Lemma}
\newtheorem{proposition}[theorem]{Proposition}
\newtheorem{corollary}[theorem]{Corollary}
\theoremstyle{definition}

\theoremstyle{remark}
\newtheorem{remark}[theorem]{Remark}

\numberwithin{equation}{section}

\begin{document}

\title{\textbf{The BCS-Bogoliubov gap equation\\
with external magnetic field\\
and the first-order phase transition}}
\author{Shuji Watanabe \\
Division of Mathematical Sciences, Graduate School of Engineering,\\ Gunma University, \\
Maebashi, 371-8510, Japan, \\
e-mail: shuwatanabe@gunma-u.ac.jp.}

\date{}

\maketitle

\begin{abstract}
\textsl{We deal with a type I superconductor in a constant external magnetic field. We obtain the BCS-Bogoliubov gap equation with external magnetic field and apply the implicit function theorem to it. We show that there is a unique magnetic field (the critical magnetic field) given by a smooth function of the temperature and that there is also a unique nonnegative solution (the gap function) given by a smooth function of both the temperature and the external magnetic field. Using the grand potential, we show that the transition from the normal state to the superconducting state in a type I superconductor is of the first order. Moreover we obtain the explicit expression for the entropy gap.} 

\medskip


\medskip

\end{abstract}

\section{Introduction}

As for as the present author knows, no one shows that the critical magnetic field is a smooth function of the temperature. Moreover, no one shows that the solution (the gap function) to the BCS-Bogoliubov gap equation with external magnetic field is a smooth function of both the temperature and the external magnetic field. In this paper we solve these problems.

To this end, we deal with a type I superconductor in the BCS-Bogoliubov model (see \cite{bcs}, \cite{bogoliubov}). Here we have a constant external magnetic field. We first consider the BCS Hamiltonian with external magnetic field and transform the mean field BCS Hamiltonian by the Bogoliubov transformation. We then obtain the BCS-Bogoliubov gap equation with external magnetic field We next apply the implicit function theorem to the BCS-Bogoliubov gap equation with external magnetic field. We show that there is a unique magnetic field (the critical magnetic field). Here, the critical magnetic field is given as the implicit function defined by the BCS-Bogoliubov gap equation with external magnetic field. We show that the critical magnetic field is a smooth function of the temperature and study its several properties.
 
On the basis of the existence of the critical magnetic field, we again apply the implicit function theorem to the BCS-Bogoliubov gap equation with external magnetic field. We show that there is a unique nonnegative solution (the gap function). Here, the gap function is again given as the implicit function defined by the BCS-Bogoliubov gap equation with external magnetic field. We show that the gap function is a smooth function of both the temperature and the external magnetic field and study its several properties.

Using the grand potential that might include the effect similar to that of the FFLO state (see \cite{fulde-ferrell}, \cite{larkin-ovchinnikov}), we show that the transition from the normal state to the superconducting state in a type I superconductor is of the first order. Moreover we obtain the explicit expression for the entropy gap. See Kashima \cite{kashima-two}--\cite{kashima-one} for the study of the BCS-Bogoliubov model with external imaginary magnetic field.

When there is no external magnetic field, the BCS-Bogoliubov gap equation with external magnetic field is reduced to the usual BCS-Bogoliubov gap equation without external magnetic field (see \cite{bcs}, \cite{bogoliubov}). There are many literatures related to the study of the BCS-Bogoliubov gap equation without magnetic field for a constant potential and for a function potential. See, e.g., \cite{bcs}, \cite{bogoliubov}, \cite{odeh}--\cite{watanabe-nine}. 
Note that the temperature is fixed in most of these literatures. As is well known, it is important to study how physical quantities such as the energy, entropy and specific heat change with changing temperature in condensed matter physics. Therefore, it is highly desirable that one does not fix the temperature, and it is highly desirable to use a function space consisting of functions of the temperature and other variables such as the energy.

In fact, using the Banach space consisting of continuous functions of both the temperature and the energy of an electron, the present author \cite{watanabe-nine} gave a new operator-theoretical proof of the statement that there is a unique nonnegative solution to the usual BCS-Bogoliubov gap equation without external magnetic field under a certain weak and simple condition. In \cite{watanabe-nine}, the potential in the BCS-Bogoliubov gap equation without external magnetic field is a function and need not be a constant. Moreover, the present author \cite{watanabe-nine} pointed out several properties of the solution such as continuity and smoothness with respect to both the temperature and the energy of an electron, and showed that the solution (the energy gap) increases monotonically as the temperature goes to zero temperature from the viewpoint of operator theory, which are in good agreement with experiments. Furthermore, the present author \cite{watanabe-nine} gave an operator-theoretical proof of the statement that the transition from the normal state to the superconducting state is of the second order without external magnetic field. We emphasize again that the potential in the BCS-Bogoliubov gap equation without external magnetic field is a function and need not be a constant.

Throughout this paper we use the unit where the Boltzmann constant $k_B$ is equal to 1.

\section{The mean field BCS Hamiltonian with external magnetic field}

We consider a type I superconductor in a constant external magnetic field $H \bm{e}_z$ along the \textit{z}-axis. Here, $H>0$ is a positive constant.

Let the vector potential $\bm{A}$ be $\bm{A}=\frac{\, H \,}{2}(-y,\, x,\, 0)$. Then, $\bm{\nabla} \times \bm{A}=H \bm{e}_z$ and $\bm{\nabla} \cdot \bm{A}=0$. We denote the electron field and its Hermitian conjugate by
\[
\Psi(\bm{r})=\sum_{k,\, \sigma} \phi_{k \sigma}(\bm{r}) \, C_{k\sigma}, \quad
\Psi ^{\dagger} (\bm{r})=\sum_{k,\, \sigma} \phi_{k \sigma}^*(\bm{r}) \, C_{k\sigma}^{\dagger},
\]
respectively. Here, $k \in \mathbb{R}^3$ denotes the wave vector, and $\sigma= \uparrow, \, \downarrow$. The set $\left\{ \phi_{k \sigma}(\bm{r})  \right\}_{k, \, \sigma}$ is a complete orthonormal set, and $C_{k\sigma}^{\dagger}$ and $C_{k\sigma}$ are creation and annihilation operators of an electron, respectively. As is well known, the kinetic energy term becomes
\begin{eqnarray*}
& &\int \Psi ^{\dagger} (\bm{r})
\frac{1}{\, 2m\,}\left( -i \hslash \bm{\nabla}+\frac{\, e\,}{c}\bm{A} \right)^2
\Psi(\bm{r}) \, d \bm{r} \\
&=& \sum_{k,k', \sigma} C_{k\sigma}^{\dagger} C_{k' \sigma}
\int \phi_{k \sigma}^* (\bm{r})
\left( -\frac{\hslash^2}{\, 2m\,} \bm{\nabla}^2+\frac{\, e H}{\, 2mc \,} 
\left( \bm{r} \times \bm{p} \right)_z+\frac{ \, e^2 H^2}{\, 8mc^2 \,} (x^2+y^2) \right)
\phi_{k' \sigma}(\bm{r}) \, d \bm{r}. 
\end{eqnarray*}
Here, $m$ is the mass of an electron with charge $-e$, and $c$ is the speed of light.

Let us suppose the following very simple approximations.
\begin{eqnarray}
\int \phi_{k \sigma}^* (\bm{r})
\left( -\frac{\hslash^2}{\, 2m\,} \bm{\nabla}^2 \right)
\phi_{k' \sigma}(\bm{r}) \, d \bm{r}
&\approx& 
\frac{\, \hslash^2k^2\,}{\, 2m\,} \delta_{k, \, k'}, \label{eqn:ap-one} \\
\int \phi_{k \sigma}^*(\bm{r}) \frac{\, e \,}{\, 2mc \,} 
\left( \bm{r} \times \bm{p} \right)_z \phi_{k' \sigma}(\bm{r}) \, d\bm{r}
&\approx&
a \, \delta_{k, \, k'}, \label{eqn:ap-two} \\
\int \phi_{k \sigma}^*(\bm{r}) \frac{ \, e^2 \,}{\, 8mc^2 \,} (x^2+y^2)
\phi_{k' \sigma}(\bm{r}) \, d\bm{r}
&\approx&
b \, \delta_{k, \, k'}, \label{eqn:ap-three}
\end{eqnarray}
where $a, \,b>0$ are positive constants. Generally speaking, it is probable that there are terms where $k \not= k'$. Moreover, it is also probable that $a$ and $b$ are both functions of the wave vectors $k$ and $k'$. But we do not take those into account here. This is because we are trying to include the effect of the external magnetic field and to construct a very simple model that leads to the first-order phase transition. We will study the case where there are terms where $k \not= k'$ and $a$ and $b$ are both functions of the wave vectors $k$ and $k'$ in an upcoming paper.

Note that $a>0$, since electrons move so as to cancel the external magnetic field $H \bm{e}_z$ and show perfect diamagnetism. See also \eqref{eqn:slope} below, where the slope of the critical magnetic field at the transition temperature implies $a>0$.

As in the BCS-Bogoliubov model (see \cite{bcs}, \cite{bogoliubov}) without external magnetic field, under the approximations above, the Hamiltonian $H$ minus the chemical potential $\mu$ times the electron number operator $N$ turns out to be
\begin{eqnarray*}
H-\mu N
&=&
\sum_{k} \left[  \left\{ \xi_k+aH+bH^2+\mu_BH  \right\} C_{k\uparrow}^{\dagger} C_{k\uparrow}+\left\{ \xi_k+aH+bH^2-\mu_BH  \right\} C_{-k\downarrow}^{\dagger} C_{-k\downarrow}
 \right] \\
& & +\sum_{k, \, k'} U_{k,\, k'} C_{k'\uparrow}^{\dagger} C_{-k'\downarrow}^{\dagger} C_{-k\downarrow} C_{k\uparrow},
\end{eqnarray*}
where $\xi_k=( \hbar^2k^2 / 2m )-\mu$ and $\mu_B$ is the Bohr magneton. Here, $U_{k,\, k'}$ is the Fourier transform of the interacting potential, but we call $U_{k,\, k'}$ the potential for simplicity.

As in the BCS-Bogoliubov model without external magnetic field, we use the mean field approximation and set
\begin{equation}\label{eqn:gapeq}
\Delta_k(T, \, H)=-\sum_{k'} U_{k,\, k'} \langle C_{-k'\downarrow} C_{k' \uparrow} \rangle_{\beta},
\end{equation}
where $\beta=1/T$ with $T \geq 0$ the absolute temperature. Set
\begin{eqnarray*}
u_k
&=&
\frac{1}{\,\sqrt{2}\,} 
 \sqrt{ 1+\frac{\xi_k+aH+bH^2}{\,  \sqrt{(\xi_k+aH+bH^2)^2+\Delta_k(T, \, H)^2 } \,} }, \\
& & \\
v_k
&=&
\frac{1}{\,\sqrt{2}\,} 
 \sqrt{ 1-\frac{\xi_k+aH+bH^2}{\,  \sqrt{(\xi_k+aH+bH^2)^2+\Delta_k(T, \, H)^2 } \,} },
\end{eqnarray*}
where $u_k$ and $v_k$ are real, and $0 \leq u_k \leq 1$ and $0 \leq v_k \leq 1$. Note that they satisfy $u_k^2+v_k^2=1$.

We use the following Bogoliubov transformation:
\begin{eqnarray*}
C_{k\uparrow}
&=&  u_k \,\gamma_{k\uparrow}+v_k \,\gamma_{-k\downarrow}^{\dagger}, \\
C_{-k\downarrow}
&=& u_k \,\gamma_{-k\downarrow}-v_k \,\gamma_{k\uparrow}^{\dagger}.
\end{eqnarray*}
The mean field BCS Hamiltonian $(H-\mu N)_{MF}$ turns out to be
\begin{eqnarray*}
(H-\mu N)_{MF}
&=&
\sum_{k} \left[  \left\{ E_k(T,\, H)+\mu_BH  \right\}
 \gamma_{k\uparrow}^{\dagger} \gamma_{k\uparrow} +\left\{ E_k(T,\, H)-\mu_BH  \right\}
 \gamma_{-k\downarrow}^{\dagger} \gamma_{-k\downarrow}\right. \\
& & \quad + \xi_k+aH+bH^2-E_k(T,\, H) \\
& & \quad +\left. \frac{\Delta_k(T, \, H)^2}{\, 2E_k(T,\, H)  \,} \,
\left( 1-\frac{1}{\, e^{\beta \{ E_k(T,\, H)+\mu_BH \}}+1  \,} -\frac{1}{\, e^{\beta \{ E_k(T,\, H)-\mu_BH \}}+1  \,} \right) \right],
\end{eqnarray*}
where $E_k(T,\, H)=\sqrt{(\xi_k+aH+bH^2)^2+\Delta_k(T, \, H)^2  }$. Here, we assume $ \langle \gamma_{-k\downarrow} \gamma_{k\uparrow}\rangle_{\beta} = \langle \gamma_{k\uparrow}^{\dagger} \gamma_{-k\downarrow}^{\dagger} \rangle_{\beta} =0$ as in the BCS-Bogoliubov model without external magnetic field. The equation \eqref{eqn:gapeq} becomes
\begin{equation}\label{eqn:gapeqtwo}
\Delta_k(T, \, H)=-\sum_{k'} \frac{\, U_{k,\, k'} \, \Delta_{k'}(T, \, H) \,}{2E_{k'}(T,\, H)} \frac{\sinh \left( \beta E_{k'}(T,\, H) \right) }{\,  \cosh \left( \beta E_{k'}(T,\, H) \right)+\cosh \left( \beta \mu_B H \right) \,}.
\end{equation}
We thus obtain the BCS-Bogoliubov gap equation \eqref{eqn:gapeqtwo}  with external magnetic field. Note that
\[
\frac{\sinh \left( \beta E_k(T,\, H) \right) }{\,  \cosh \left( \beta E_k(T,\, H) \right)+\cosh \left( \beta \mu_B H \right) \,}
=1-\frac{1}{\, e^{\beta \{ E_k(T,\, H)+\mu_BH \}}+1  \,} -\frac{1}{\, e^{\beta \{ E_k(T,\, H)-\mu_BH \}}+1  \,}. 
\]

\section{The grand potential with external magnetic field}

The grand potential is a function of the temperature, the volume of the system and the chemical potential. However, we fix both the volume and the chemical potential. Moreover, we have the external magnetic field $H$. Therefore, in our case, the grand potential depends on both the temperature and the external magnetic field. So we denote it by $\Omega(T, \, H)$.

The grand potential $\Omega(T, \, H)$ with external magnetic field is given by the partition function $Z={\rm Tr} ( e^{-\beta \, (H-\mu N)_{MF}} )$ and is of the form
\begin{eqnarray}\label{eqn:gpo}
& & \Omega(T, \, H) \\
&=&
\frac{1}{\, 2\,} \sum_{k \in \mathbb{R}^3,\, \sigma=\uparrow} \left[ \xi_k+aH+bH^2
-\frac{\, \left( \xi_k+aH+bH^2 \right)^2 \,}{E_k(T,\, H)}   \right. 
\nonumber \\
& &
\hspace{20mm} -\left. \frac{\, \Delta_k(T, \, H)^2 \,}{\, E_k(T,\, H)  \,} \,
\frac{1}{\, 1+e^{\beta (E_k(T,\, H)+\mu_BH)}  \,} 
-2T \ln \left(  1+e^{-\beta (E_k(T,\, H)+\mu_BH)} \right) \right] 
\nonumber \\
& & \nonumber \\
& &
\hspace{-3mm} +\frac{1}{\, 2\,} \sum_{k \in \mathbb{R}^3,\, \sigma=\downarrow} \left[   \xi_k+aH+bH^2
 -\frac{\, \left( \xi_k+aH+bH^2 \right)^2+2\Delta_k(T, \, H)^2 \,}{E_k(T,\, H)}  \right. \nonumber \\
& &
\hspace{20mm} +\left. 
\frac{\, \Delta_k(T, \, H)^2 \,}{\, E_k(T,\, H)  \,} \,
\frac{1}{\, 1+e^{-\beta (E_k(T,\, H)-\mu_B H)}  \,} 
-2T \ln \left(  1+e^{-\beta (E_k(T,\, H)-\mu_B H)} \right) \right]. 
\nonumber
\end{eqnarray}
The first term on the right corresponds to the contribution from particles with spin up, and the second term from particles with spin down.

In order to deal with the grand potential from the viewpoint of mathematical analysis, we replace the summation over the wave vector in \eqref{eqn:gpo} by
the integral with respect to the wave vector. Here the integral domain is a certain bounded and closed subset of the wave vector space $\mathbb{R}^3$. On the other hand, if the potential in the BCS-Bogoliubov gap equation \eqref{eqn:gapeqtwo} is a function of $k^2$ and $k'^2$, not a function of $k$ and $k'$, then we replace the summation over the wave vector in \eqref{eqn:gpo} by the integral with respect to the squared wave vector. Here the integral domain is also a certain bounded and closed interval of the squared wave vector space $\mathbb{R}$. Therefore we can deal with both cases similarly, and so we assume that the potential is a function of $k^2$ and $k'^2$. Then the potential turns out to be a function of $\xi=(\hslash^2k^2/2m)-\mu$ and $\xi'=(\hslash^2k'^2/2m)-\mu$. Therefore we denote the potential by $U(\xi,\, \xi')$. Accordingly, the solution (the gap function) to the BCS-Bogoliubov gap equation becomes a function of $T$, $H$ and $\xi$. We denote it by $\Delta_{\xi}(T, \, H)$ and set
\[
E_{\xi}(T,\, H)=\sqrt{(\xi+aH+bH^2)^2+\Delta_{\xi}(T, \, H)^2 }.
\]

By the approximations \eqref{eqn:ap-one}, \eqref{eqn:ap-two} and \eqref{eqn:ap-three}, the energy $\varepsilon$ of a particle with spin up is given by
\begin{equation}\label{eqn:up}
\varepsilon=\frac{\, \hslash^2k^2\,}{\, 2m\,}+aH+bH^2+\mu_B H
=\xi+\mu+aH+bH^2+\mu_B H,
\end{equation}
while that of a particle with spin down is given by
\begin{equation}\label{eqn:down}
\varepsilon=\frac{\, \hslash^2k^2\,}{\, 2m\,}+aH+bH^2-\mu_B H
=\xi+\mu+aH+bH^2-\mu_B H.
\end{equation}

In the form of the grand potential \eqref{eqn:gpo}, we replace the summation over $k$ by the integral over the energy $\varepsilon$ from $\varepsilon_F-\hslash \omega_D$ to $\varepsilon_F+\hslash \omega_D$. Here, $\varepsilon_F=\mu$. See figure 1.

\begin{figure}[htbp]
\includegraphics[width=15cm]{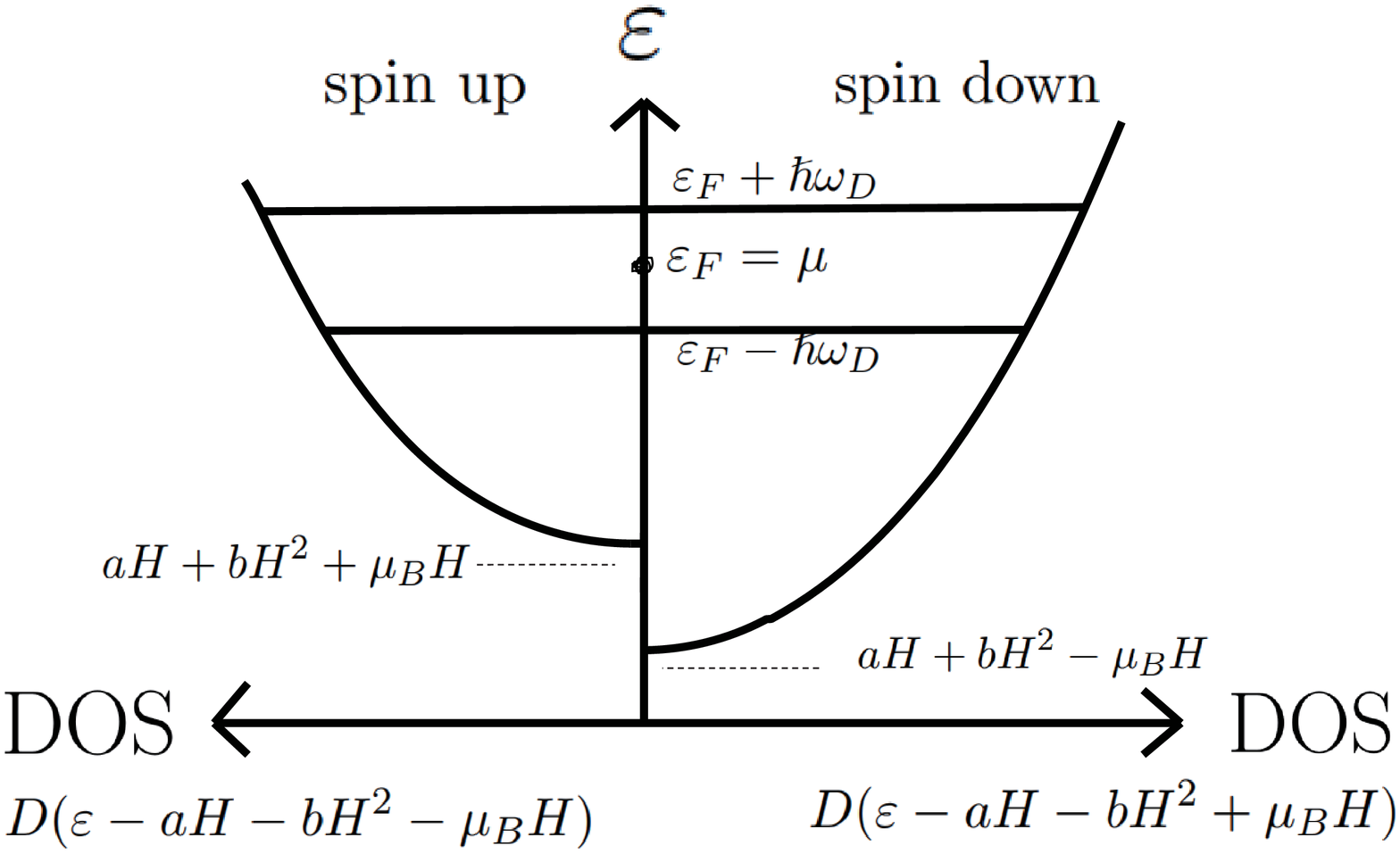}
\caption{\textsf{Density of states (DOS).}}
\end{figure}

\begin{remark} \quad The wave vector $k_{F \uparrow}$ of a particle with spin up corresponding to $\varepsilon_F$ satisfies
\[
\frac{\, \hslash^2k_{F \uparrow}^2\,}{\, 2m\,}
=\mu-aH-bH^2-\mu_B H,
\]
while the wave vector $k_{F \downarrow}$ of a particle with spin down corresponding to $\varepsilon_F$ satisfies
\[
\frac{\, \hslash^2k_{F \downarrow}^2\,}{\, 2m\,}
=\mu-aH-bH^2+\mu_B H.
\]
Therefore,
\[
k_{F \uparrow}^2<k_{F \downarrow}^2.
\]
See figure 1.
\end{remark}

Then
\begin{eqnarray}\label{eqn:gpotwo}
& & \\
& & \Omega(T, \, H) \nonumber \\
&=&
\frac{1}{\, 2\,} \int_{\varepsilon_F-\hslash \omega_D}^{\varepsilon_F+\hslash \omega_D} d\varepsilon \,
D(\varepsilon-aH-bH^2-\mu_B H) \left[ \xi+aH+bH^2
-\frac{\, \left( \xi+aH+bH^2 \right)^2 \,}{E_{\xi}(T,\, H)}   \right. 
\nonumber \\
& &
\hspace{20mm} -\left. 
\frac{\, \Delta_{\xi}(T, \, H)^2 \,}{\, E_{\xi}(T,\, H)  \,} \,
\frac{1}{\, 1+e^{\beta (E_{\xi}(T,\, H)+\mu_BH)}  \,} 
-2T \ln \left(  1+e^{-\beta (E_{\xi}(T,\, H)+\mu_BH)} \right) \right] 
\nonumber \\
& & \nonumber \\
& &
\hspace{-3mm} +\frac{1}{2}\int_{\varepsilon_F-\hslash \omega_D}^{\varepsilon_F+\hslash \omega_D} d\varepsilon \,
D(\varepsilon-aH-bH^2+\mu_B H)
\left[   \xi+aH+bH^2
 \phantom{-\frac{\, \left( \xi+aH+bH^2 \right)^2+2\Delta_{\xi}(T, \, H)^2 \,}{E_{\xi}(T,\, H)}} 
\right. \nonumber \\
& &
\hspace{20mm} \left. -\frac{\, \left( \xi+aH+bH^2 \right)^2+2\Delta_{\xi}(T, \, H)^2 \,}{E_{\xi}(T,\, H)}  \right. \nonumber \\
& &
\hspace{20mm} +\left. 
\frac{\, \Delta_{\xi}(T, \, H)^2 \,}{\, E_{\xi}(T,\, H)  \,} \,
\frac{1}{\, 1+e^{-\beta (E_{\xi}(T,\, H)-\mu_B H)}  \,} 
-2T \ln \left(  1+e^{-\beta (E_{\xi}(T,\, H)-\mu_B H)} \right) \right],
\nonumber
\end{eqnarray}
where $\omega_D$ denotes the Debye angular frequency and $D(\cdot)$ the density of states.

\begin{remark} \label{rmk:FFLO} \quad Let $k_{min \uparrow}$ be the wave vector of a particle with spin up corresponding to $\varepsilon_F-\hslash \omega_D$. Then each wave vector $k_{\downarrow}$ of a particle with spin down satisfies
\[
k_{min \uparrow}^2 < k_{\downarrow}^2,
\]
provided $H \not= 0$. See figure 1. Therefore, there is no $k_{\downarrow}^2$ that is exactly equal to $k_{min \uparrow}^2$. On the other hand, let $k_{max \downarrow}$ be the wave vector of a particle with spin down corresponding to $\varepsilon_F+\hslash \omega_D$. Then each wave vector $k_{\uparrow}$ of a particle with spin up satisfies
\[
k_{max \downarrow}^2 > k_{\uparrow}^2,
\]
provided $H \not= 0$. Therefore, there is no $k_{\uparrow}^2$ that is exactly equal to $k_{max \downarrow}^2$. So we take into account the effect that \textit{might} be similar to that of the FFLO state in the form of the grand potential \eqref{eqn:gpotwo}.
\end{remark}

Taking \eqref{eqn:up} and \eqref{eqn:down} into account, we change the variable $\varepsilon$ into the new variable $\xi$ to obtain
\begin{eqnarray}\label{eqn:gpothree}
& & \\
& & \Omega(T, \, H) \nonumber \\
&=&
\frac{1}{\, 2\,} \int_{I_{\uparrow}} d\xi \,
D(\xi+\mu) \left[ \xi+aH+bH^2
-\frac{\, \left( \xi+aH+bH^2 \right)^2 \,}{E_{\xi}(T,\, H)}   \right. 
\nonumber \\
& &
\hspace{20mm} -\left. 
\frac{\, \Delta_{\xi}(T, \, H)^2 \,}{\, E_{\xi}(T,\, H)  \,} \,
\frac{1}{\, 1+e^{\beta (E_{\xi}(T,\, H)+\mu_BH)}  \,} 
-2T \ln \left(  1+e^{-\beta (E_{\xi}(T,\, H)+\mu_BH)} \right) \right] 
\nonumber \\
& & \nonumber \\
& &
\hspace{-3mm} +\frac{1}{2}\int_{I_{\downarrow}} d\xi \,
D(\xi+\mu) \left[   \xi+aH+bH^2
-\frac{\, \left( \xi+aH+bH^2 \right)^2+2\Delta_{\xi}(T, \, H)^2 \,}{E_{\xi}(T,\, H)}  \right. \nonumber \\
& &
\hspace{20mm} +\left. 
\frac{\, \Delta_{\xi}(T, \, H)^2 \,}{\, E_{\xi}(T,\, H)  \,} \,
\frac{1}{\, 1+e^{-\beta (E_{\xi}(T,\, H)-\mu_B H)}  \,} 
-2T \ln \left(  1+e^{-\beta (E_{\xi}(T,\, H)-\mu_B H)} \right) \right],
\nonumber
\end{eqnarray}
where
\begin{eqnarray*}
I_{\uparrow}
&=&
[-\hslash\omega_D-(aH+bH^2)-\mu_BH,
 \, \hslash\omega_D-(aH+bH^2)-\mu_BH], \\
I_{\downarrow}
&=&
[-\hslash\omega_D-(aH+bH^2)+\mu_BH,
 \, \hslash\omega_D-(aH+bH^2)+\mu_BH].
\end{eqnarray*}

We define the grand potential for the normal state by
\[
\Omega_N(T, \, H)=\Omega(T, \, H) |_{\Delta_{\xi}=0},
\]
and that for the superconducting state by
\[
\Omega_S(T, \, H)=\Omega(T, \, H).
\]
We mainly deal with the difference between the two above:
\begin{equation}\label{eqn:psi}
\Psi(T, \, H)=\Omega_S(T, \, H)-\Omega_N(T, \, H).
\end{equation}

\section{The BCS-Bogoliubov gap equation with external magnetic field}

The grand potential $\Omega(T, \, H)$ has the gap function $\Delta_{\xi}(T, \, H)$ in its form, as is shown in \eqref{eqn:gpothree}. So we have to show that there is a unique solution (the gap function) to the BCS-Bogoliubov gap equation with external magnetic field so as to give meaning to the grand potential. Note that we use the unit where the Boltzmann constant $k_B$ is equal to 1, as mentioned in the introduction.

In this section we therefore show the existence of the gap function and study its properties such as continuity, smoothness and monotone decreasingness with respect to both the temperature $T$ and the external magnetic field $H$. Since we are trying to include the effect of the external magnetic field and to construct a very simple model that leads to the first-order phase transition, we approximate each of the interval $I_{\uparrow}$ and $I_{\downarrow}$ by $I=[-\hslash\omega_D, \, \hslash\omega_D]$, i.e., we put
\[
I_{\uparrow}=I_{\downarrow}=[-\hslash\omega_D, \, \hslash\omega_D]=I.
\]
Actually, $\displaystyle{ I_{\uparrow} \not=I }$ and $\displaystyle{  I_{\downarrow} \not= I }$, and so there is the contribution of the fact that $\displaystyle{ I_{\uparrow} \not=I }$ and $\displaystyle{  I_{\downarrow} \not= I }$ to the gap function. We will study the case where $\displaystyle{ I_{\uparrow} \not=I }$ and $\displaystyle{  I_{\downarrow} \not= I }$ in an upcoming paper.

Under this approximation, the BCS-Bogoliubov gap equation \eqref{eqn:gapeqtwo} with external magnetic field becomes
\begin{equation}\label{eqn:gapfour}
\Delta_x(T, \, H)= -\int_I 
\frac{\,  D(\xi+\mu) \, U(x,\, \xi) \, \Delta_{\xi}(T, \, H) \,}{\, 2E_{\xi}(T,\, H)  \,}  \frac{\sinh \left( \beta E_{\xi}(T,\, H) \right) }{\,  \cosh \left( \beta E_{\xi}(T,\, H) \right)+\cosh \left( \beta \mu_B H \right) \,} \, d\xi,
\end{equation}
where
\[
E_{\xi}(T,\, H)=\sqrt{(\xi+aH+bH^2)^2+\Delta_{\xi}(T, \, H)^2 }.
\]
Here we replace $U(\xi,\, \xi')$ by $U(x,\, \xi)$.

As in the BCS-Bogoliubov model without external magnetic field, we deal with the case where $U_1$ is a positive constant. Here,
\begin{equation}\label{eqn:constant}
U_1=-\frac{\,  D(\xi+\mu) \, U(x,\, \xi) \,}{\, 2 \,} \; \left( >0 \right).
\end{equation}
We will study the case where $U_1$ is not a constant, i.e., $D(\xi+\mu) \, U(x,\, \xi)$ is a function of $x$ and $\xi$ in an upcoming paper, as mentioned in the last section.

Since $U_1$ is a constant, the gap function $\Delta_x(T, \, H)$ does not depend on $x$. So we denote it by $\Delta(T, \, H)$. Therefore, the BCS-Bogoliubov gap equation \eqref{eqn:gapfour} turns out to be
\begin{equation}\label{eqn:gapfive}
\int_I \frac{\, 1 \,}{\, E_{\xi}(T,\, H)  \,}  \frac{\sinh \left( \beta E_{\xi}(T,\, H) \right) }{\,  \cosh \left( \beta E_{\xi}(T,\, H) \right)+\cosh \left( \beta \mu_B H \right) \,} \, d\xi=\frac{1}{\, U_1 \,}.
\end{equation}

We set
\[
Y=\Delta(T, \, H)^2
\]
and define
\begin{equation}\label{eqn:function-F}
F(T, \, H,\, Y)=
\int_I \frac{1}{\, E_{\xi}(H, \, Y)  \,} \, 
\frac{\sinh \left( \beta E_{\xi}(H, \, Y) \right) }{\,  \cosh \left( \beta E_{\xi}(H, \, Y) \right)+\cosh \left( \beta \mu_B H \right) \,}
 \, d\xi-\frac{1}{\, U_1\,},
\end{equation}
where
\[
E_{\xi}(H,\, Y)=\sqrt{(\xi+aH+bH^2)^2+Y }.
\]

\begin{remark}
The symbol $F$ does not denote a Legendre transformation here. It just denotes a function of the three variables $T$, $H$ and $Y$. Its physical meaning is that the BCS-Bogoliubov gap equation \eqref {eqn:gapfive} with external magnetic field is rewritten by
\[
F(T, \, H,\, \Delta(T, \, H)^2)=0.
\]
\end{remark}

First of all, let us define the transition temperature (the critical temperature) $\tau_1$ without external magnetic field:
\begin{eqnarray}\label{eqn:tau-one}
F(\tau_1, \, 0,\, 0)
&=&
\int_I \frac{d\xi}{\, |\xi|  \,} \frac{\sinh \left( |\xi|/\tau_1 \right) }{\,  \cosh \left(  |\xi|/\tau_1 \right)+\cosh \left( 0  \right) \,}-\frac{1}{\, U_1\,} \\
&=&
\int_I \frac{d\xi}{\, \xi  \,} \tanh \frac{ \xi }{\, 2\tau_1 \,}-\frac{1}{\, U_1\,}=0. \nonumber
\end{eqnarray}
This means that the gap function is equal to $0$ when $T=\tau_1$ and $H=0$, i.e., $(\sqrt{Y}=) \, \Delta(\tau_1, \, 0)=0$. Note that $\tau_1$ is the transition temperature without external magnetic field.

Let $T_0$ be arbitrary as long as $0<T_0<\tau_1$. Let $Y_0>0$ be arbitrary but large enough. Set
\begin{equation}\label{eqn:domain-d}
D=[T_0,\, \tau_1] \times \left[ 0,\, \frac{\, 1.24 \, T_0 \,}{\mu_B} \right] \times [0,\, Y_0].
\end{equation}
We define the function $F$ (see \eqref{eqn:function-F}) on the set $D$.

\begin{remark}\label{rmk:domaind}
If $H \leq 1.24 \, T_0/\mu_B$, then a straightforward calculation gives
\[
\cosh \frac{\, \mu_B H \,}{T} \left( \sinh z-z\cosh z  \right)+\cosh z \sinh z-z \geq 0,
\]
where $z=E_{\xi}(H,\, Y)/T$. Therefore,
\[
\frac{\partial F}{\, \partial H \,}(T, \, H,\, Y)<0.
\]
which is necessary to show (2) of the following lemma. So we define the domain $D$ as in \eqref{eqn:domain-d}.
\end{remark}

\begin{lemma}\label{lm:properties of F} \quad Let $D$ be as in \eqref{eqn:domain-d}.

\noindent \rm{(1)} \quad The function F is uniformly continuous on $D$.

\noindent \rm{(2)} \quad $F \in C^1(D)$. Moreover, $\displaystyle{  \frac{\partial F}{\, \partial H \,}(T, \, H,\, Y)<0 }$, and hence $F$ is strictly decreasing with respect to $H$.

\noindent \rm{(3)} \quad Let $T_0 \le T <\tau_1$. Then \quad $\displaystyle{
F(T, \, 0,\, 0)>0 }$.
\end{lemma}

\begin{proof} \rm{(1)} \quad We set
\[
F(T, \, H,\, Y)=\int_I J(T,\, H, \, Y,\, \xi) \, d\xi-\frac{1}{\, U_1\,},
\]
where
\[
J(T,\, H, \, Y,\, \xi)= \frac{1}{\, E_{\xi}(H, \, Y)  \,} \, 
\frac{\sinh \left( E_{\xi}(H, \, Y)/T \right) }{\,  \cosh \left( 
E_{\xi}(H, \, Y)/T \right)+\cosh \left( \mu_B H/T \right) \,}.
\]
Let $(T, \, H,\, Y), \; (T_1, \, H_1,\, Y_1) \in D$. It then follows
\begin{eqnarray*}
& & J(T,\, H, \, Y,\, \xi)-J(T_1, \, H_1,\, Y_1,\,\xi) \\
&=&
(T-T_1)\frac{\partial J}{\, \partial T \,}(T_2, \, H_2,\, Y_2,\,\xi)+
(H-H_1)\frac{\partial J}{\, \partial H \,}(T_2, \, H_2,\, Y_2,\,\xi)+
(Y-Y_1)\frac{\partial J}{\, \partial Y \,}(T_2, \, H_2,\, Y_2,\,\xi),
\end{eqnarray*}
where $(T_2, \, H_2,\, Y_2,\,\xi) \in D \times I$. A straightforward calculation gives
\[
\left| J(T,\, H, \, Y,\, \xi)-J(T_1, \, H_1,\, Y_1,\,\xi) \right| \leq
M_1 | T-T_1 |+M_2 | H-H_1 |+M_3 | Y-Y_1 |,
\]
where $M_1$, $M_2$ and $M_3$ are positive constants. The constant $M_1$ depends neither on $(T,\, H, \, Y,\, \xi)$ nor on $(T_1, \, H_1,\, Y_1,\,\xi)$. The same is true for $M_2$ and $M_3$. Therefore, for an arbitrary $\varepsilon>0$, there is a $\delta>0$ such that
\begin{eqnarray*}
\left| F(T,\, H, \, Y)-F(T_1, \, H_1,\, Y_1) \right|
&\leq&
2\hslash\omega_D \, \max(M_1,\, M_2,\, M_3) \, \left( | T-T_1 |+| H-H_1 |+| Y-Y_1 | \right) \\
&<& \varepsilon,
\end{eqnarray*}
where
\[
| T-T_1 |+| H-H_1 |+| Y-Y_1 |<\delta=
\frac{\varepsilon}{\, 2\hslash\omega_D \max(M_1,\, M_2,\, M_3) \,}.
\]
Since $\delta$ depends neither on $(T,\, H, \, Y)$ nor on $(T_1, \, H_1,\, Y_1)$, the function F is uniformly continuous on $D$.

\noindent \rm{(2)} \quad A proof similar to that in (1) gives $F \in C^1(D)$. Since $H \leq 1.24 \, T_0/\mu_B$ (see \eqref{eqn:domain-d} and Remark \ref{rmk:domaind}), it follows that
\[
\cosh \frac{\, \mu_B H \,}{T} \left( \sinh z-z\cosh z  \right)+\cosh z \sinh z-z \geq 0,
\]
where $z=E_{\xi}(H,\, Y)/T$. Therefore,
\[
\frac{\partial F}{\, \partial H \,}(T, \, H,\, Y)<0.
\]

\noindent \rm{(3)} \quad A straightforward calculation gives
\[
F(T, \, 0,\, 0)=
\int_I \frac{d\xi}{\, \xi  \,} \tanh \frac{ \xi }{\, 2T \,}-\frac{1}{\, U_1\,}>
\int_I \frac{d\xi}{\, \xi  \,} \tanh \frac{ \xi }{\, 2\tau_1 \,}-\frac{1}{\, U_1\,}=0.
\]
\end{proof}

The following lemma shows the existence and uniqueness of the critical magnetic field $H_c(T)$ and moreover shows the smoothness of $H_c(T)$ with respect to the temperature $T$. 

\begin{lemma}\label{lm:hct}
Let $(T, \, H,\, 0) \in D$. Then the equality $F(T, \, H,\, 0)=0$ defines a unique magnetic field $H_c(T)$ (the critical magnetic field) implicitly. Therefore,  the critical magnetic field $H_c(T)$ is the implicit function defined by the equality $F(T, \, H,\, 0)=0$, and satisfies $F(T, \, H_c(T),\, 0)=0$. Moreover, if $T_0$ is chosen such that $H_c(T_0) \leq (1.24 \, T_0/\mu_B)$, then the critical magnetic field $H_c(T)$ is in $C^1[T_0,\, \tau_1]$.
\end{lemma}

\begin{proof}
\textit{Step 1}. \quad By \eqref{eqn:tau-one}, $F(\tau_1, \, 0,\, 0)=0$. From Lemma \ref{lm:properties of F} (2), it follows that the function $F$ is strictly decreasing with respect to $H$. Therefore, there is a point ${\rm P_1}(\tau_1, \, H_1,\, 0) \in D$ satisfying $F(\tau_1, \, H_1,\, 0)<0$  (see figure 2). Since $F$ is continuous on $D$ (see Lemma \ref{lm:properties of F} (1)), there is a point ${\rm P_2}(T_1, \, H_1,\, 0) \in D$ satisfying $F(T_1, \, H_1,\, 0)<0$ (see figure 2). Here, $T_0 \leq T_1<\tau_1$. Note that $F(T_1, \, 0,\, 0)>0$ (see Lemma \ref{lm:properties of F} (3)). Since $F$ is strictly decreasing with respect to $H$, there is a unique point ${\rm P_3}(T_1, \, H_c(T_1),\, 0) \in D$ satisfying $F(T_1, \, H_c(T_1),\, 0)=0$  (see figure 2).

\textit{Step 2}. \quad Since $F$ is strictly decreasing with respect to $H$, there is a point ${\rm P_4}(T_1, \, H_2,\, 0) \in D$ satisfying both $H_c(T_1)<H_2$ and $F(T_1, \, H_2,\, 0)<0$  (see figure 2). Since $F$ is continuous on $D$, there is a point ${\rm P_5}(T_2, \, H_2,\, 0) \in D$ satisfying $F(T_2, \, H_2,\, 0)<0$. Since $F(T_2, \, 0,\, 0)>0$ and $F$ is strictly decreasing with respect to $H$, there is a unique point ${\rm P_6}(T_2, \, H_c(T_2),\, 0) \in D$ satisfying $F(T_2, \, H_c(T_2),\, 0)=0$.

\textit{Step 3}. \quad A similar discussion shows that for $T \in [T_0, \, \tau_1]$, there is a unique critical magnetic field $H_c(T)$ satisfying $F(T, \, H_c(T),\, 0)=0$. Note that
\[
H_c(\tau_1)=0
\]
by the equality $F(\tau_1, \, 0,\, 0)=0$. Thus the critical magnetic field $H_c(T)$ is the implicit function defined by the equality $F(T, \, H,\, 0)=0$. Since $F \in C^1(D)$ (see Lemma \ref{lm:properties of F} (2)), it follows $H_c(T) \in C^1[T_0,\, \tau_1]$.
\end{proof}

See figure 2 for the graph of the critical magnetic field $H_c(T)$.

\begin{figure}[htbp]
\hspace{3cm}
\includegraphics[width=12cm]{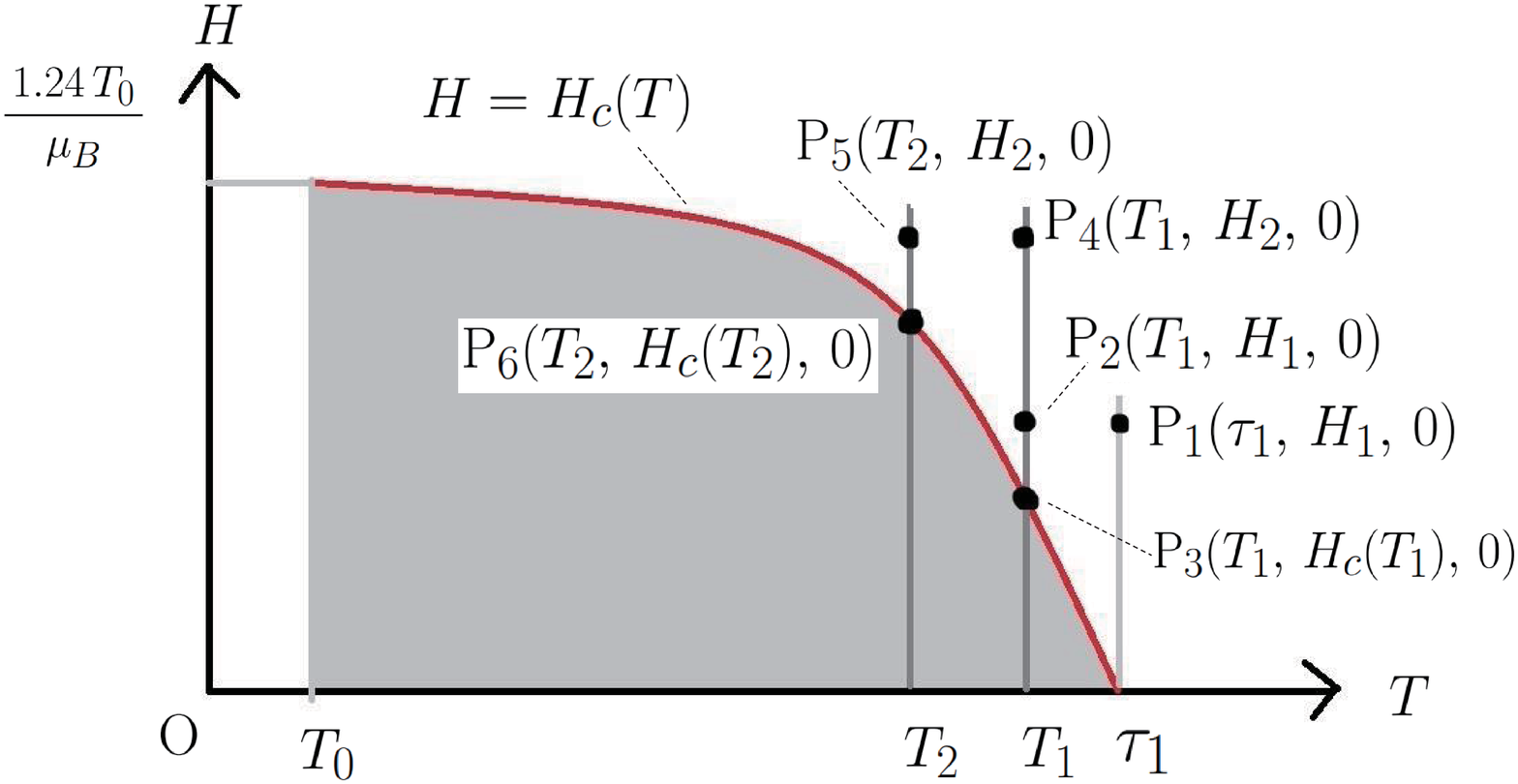}
\caption{\textsf{The critical magnetic field $H_c(T)$.}}
\end{figure}

Since $a>0$ (see \eqref{eqn:ap-two}) and $H_c(T) \in C^1[T_0,\, \tau_1]$, Lemma \ref{lm:hct} implies
\begin{equation}\label{eqn:slope}
\frac{d H_c}{\, d T \,}(\tau_1)=
-\frac{1}{\, a \, \tau_1 \,} \, \frac{ \displaystyle{ \int_I  \frac{1}{1+\cosh(\xi/\tau_1)  } \, d\xi } }
{\, \displaystyle{ \int_I  \frac{ \frac{\sinh (\xi/\tau_1)}{\xi/\tau_1}-1 }{1+\cosh(\xi/\tau_1)  } \, d\xi }  \,}<0.
\end{equation}
But, if $a=0$, then $(dH_c(\tau_1)/d T)=-\infty$, which contradicts experimental results. Moreover, if $a<0$, then $(dH_c(\tau_1)/d T)>0$, which again contradicts experimental results. Therefore we set $a>0$ in the approximation \eqref{eqn:ap-two}.

Let us show that the gap function with external magnetic field is the implicit function defined by the equality $F(T, \, H,\, Y)=0$ (the BCS-Bogoliubov gap equation).

\begin{proposition}\label{prp:f}
Let $T_0$ be as in Lemma \ref{lm:hct} and let $(T, \, H,\, Y) \in D$.

\noindent \rm{(1)} \quad Then the equality $F(T, \, H,\, Y)=0$ defines a unique nonnegative function $Y=f(T,\, H)$ $(=\Delta(T, \, H)^2)$ implicitly. Therefore, the function $Y=f(T,\, H)$ is the implicit function defined by the equality $F(T, \, H,\, Y)=0$, and satisfies $F(T, \, H,\, f(T,\, H))=0$.

\noindent \rm{(2)} \quad The domain of the function $f$ is
\[
\Omega=\left\{ \left(T,\, H \right)
\in [T_0,\, \tau_1] \times [0,\,  (1.24 \, T_0/\mu_B)] \, : \, 
0 \leq H \leq H_c(T) \right\},
\]
and $f$ satisfies
\begin{eqnarray*}
f(T, \, H)
&>&
0 \quad \hbox{at} \quad (T, \, H) \in 
\left\{ \left(T,\, H \right)
\in [T_0,\, \tau_1] \times [0,\,  (1.24 \, T_0/\mu_B)] \, : \, 
0 \leq H < H_c(T) \right\}, \\ 
f(T,\, H_c(T))
&=&
0.
\end{eqnarray*}

\noindent \rm{(3)} \quad $f \in C^1(\Omega)$. Moreover,
\[
\frac{\partial f}{\, \partial T}(T, \, H)<0, \quad 
\frac{\partial f}{\, \partial H}(T, \, H)<0 \quad \hbox{at} \quad (T, \, H) \in \Omega.
\]
\end{proposition}

\begin{proof} \quad \rm{(1)} \quad Let $(T_1, \, H_1,\, 0) \in D$ satisfy $0 \leq H_1 < H_c(T)$ (see figure 3). Then $F(T_1, \, H_1,\, 0)>0$. A straightforward calculation gives that for $(T, \, H,\, Y) \in D$,
\[
\frac{\partial F}{\, \partial Y}(T, \, H,\, Y)<0.
\]
Therefore, $F$ is strictly decreasing with respect to $Y$. Moreover,
\[
F(T, \, H,\, Y) \to -\frac{1}{\, U_1\,} < 0 \quad \hbox{as} \quad Y \to \infty.
\]
Then there is a unique $Y_1=f(T_1,\, H_1)$ $(=\Delta(T_1, \, H_1)^2)$ satisfying
\[
F(T_1, \, H_1,\, f(T_1,\, H_1))=0.
\]
See figure 3. The function $Y=f(T,\, H)$ thus obtained is the implicit function defined by $F(T, \, H,\, Y)=0$ and satisfies
\[
F(T, \, H,\, f(T,\, H))=0.
\]

\rm{(2)} follows from \rm{(1)}.

\rm{(3)} \quad Since $F \in C^1(D)$ (see Lemma \ref{lm:properties of F} (2)), it follows that $f \in C^1(\Omega)$. Note that $f(T,\, H)$ is the implicit function defined by $F(T, \, H,\, Y)=0$. A straightforward calculation then gives
\[
\frac{\partial f}{\, \partial T \,}(T,\, H)=
-\frac{\frac{\partial F}{\, \partial T}(T, \, H,\, f(T,\, H)) }
{\, \frac{\partial F}{\, \partial Y}(T, \, H,\, f(T,\, H))  \,}, \quad
\frac{\partial f}{\, \partial H \,}(T,\, H)=
-\frac{\frac{\partial F}{\, \partial H}(T, \, H,\, f(T,\, H)) }
{\, \frac{\partial F}{\, \partial Y}(T, \, H,\, f(T,\, H))  \,}.
\]
As mentioned just above,
\[
\frac{\partial F}{\, \partial Y}(T, \, H,\, Y)<0.
\]
By Lemma \ref{lm:properties of F} (2),
\[
\frac{\partial F}{\, \partial H \,}(T, \, H,\, Y)<0.
\]
Since $z_1=(\mu_B H/T) \leq 1.24$, it follows that $z_1 \sinh z_1<2$. Therefore, for $z \geq 0$,
\[
1+\cosh z \cosh z_1-z_1 \sinh z_1 \frac{\, \sinh z \,}{z}>
1+\cosh z-2\frac{\, \sinh z \,}{z} \geq 0,
\]
and hence
\[
\frac{\partial F}{\, \partial T \,}(T, \, H,\, Y)<0.
\]
Thus
\[
\frac{\partial f}{\, \partial T}(T, \, H)<0, \quad 
\frac{\partial f}{\, \partial H}(T, \, H)<0.
\]
\end{proof}

\begin{figure}[htbp]
\hspace{2cm}
\includegraphics[width=13cm]{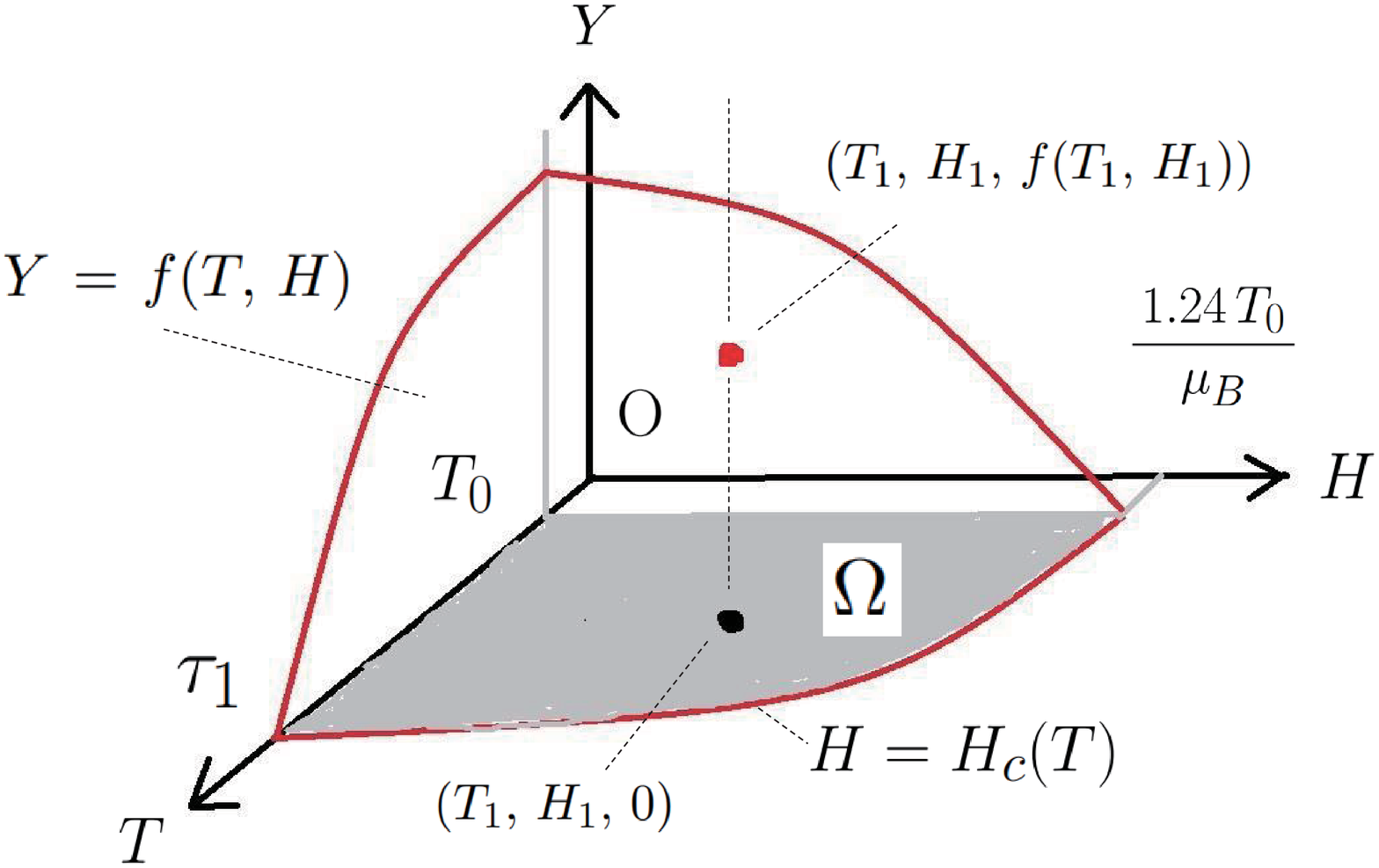}
\caption{\textsf{The graph of the squared gap function $Y=f(T,\, H)$.}}
\end{figure}

\begin{remark}
The proposition just above and the corollary just below hold true even when $H=0$, which corresponds to the BCS model without external magnetic field. See Watanabe \cite{watanabe-nine}.
\end{remark}

\bigskip

The following are several properties of the gap function $\Delta(T,\, H) \; (=\sqrt{f(T,\, H)})$ with external magnetic field. Since (see Proposition \ref{prp:f} (2))
\[
\Delta(T,\, H_c(T)) =\sqrt{f(T,\, H_c(T))} =0, \quad T \in [T_0,\, \tau_1],
\]
Proposition \ref{prp:f} immediately gives the following.

\begin{corollary}
\rm{(1)} \quad $\Delta(T, \, H) \in C([T_0,\, \tau_1] \times [0,\, (1.24 \, T_0/\mu_B)])$.

\noindent \rm{(2)} \quad $\displaystyle{
\Delta(T, \, H) > 0 \quad \hbox{at} \quad (T, \, H) \in 
\left\{ \left(T,\, H \right)
\in [T_0,\, \tau_1] \times [0,\,  (1.24 \, T_0/\mu_B)] \, : \, 
0 \leq H < H_c(T) \right\}
}$,
 
$\displaystyle{
\Delta(T,\, H_c(T))=0}$.

\noindent \rm{(3)} \quad $\Delta(T, \, H) \in C^1\left( 
\left\{ \left(T,\, H \right)
\in [T_0,\, \tau_1] \times [0,\,  (1.24 \, T_0/\mu_B)] \, : \, 
0 \leq H < H_c(T) \right\}
\right)$. Moreover,
\[
\frac{\partial \Delta}{\, \partial T}(T, \, H)<0, \quad 
\frac{\partial \Delta}{\, \partial H}(T, \, H)<0 \quad
\]
at \quad $(T, \, H) \in 
\left\{ \left(T,\, H \right)
\in [T_0,\, \tau_1] \times [0,\,  (1.24 \, T_0/\mu_B)] \, : \, 
0 \leq H < H_c(T) \right\}$.

\noindent \rm{(4)} \quad Let $(T, \, H) \in 
\left\{ \left(T,\, H \right) \in [T_0,\, \tau_1] \times [0,\,  (1.24 \, T_0/\mu_B)] \, : \, 0 \leq H < H_c(T) \right\}$. Then
\[
\frac{\partial \Delta}{\, \partial T}(T, \, H) \to -\infty, \quad
\frac{\partial \Delta}{\, \partial H}(T, \, H) \to -\infty \quad \hbox{as} \quad (T,\, H) \to (T_1,\, H_c(T_1)),
\]
where $T_1 \in [T_0,\, \tau_1]$.
\end{corollary}

\section{A proof of the first-order phase transition}

Let $\Psi$ be as in \eqref{eqn:psi}. Since $\Delta(T,\, H_c(T))=0$, it follows
\[
\Psi(T,\, H_c(T))=0.
\]
Therefore, the grand potential is continuous on $[T_0,\, \tau_1] \times [0,\, (1.24\, T_0/\mu_B)]$.

On the other hand, there is an entropy gap at $(T,\, H_c(T))$, and the entropy gap $\Delta S$ is explicitly given by
\begin{eqnarray*}
\Delta S
&=&
-\frac{\partial \Psi}{\, \partial T \,}(T,\, H_c(T)) \\
&=&
-\frac{1}{4} \frac{\partial f}{\, \partial T \,}(T, \, H_c(T))
\left\{ \int_{I_1} 
\frac{ D(\xi+\mu)  }{|\xi+aH_c(T)+bH_c(T)^2|} \, d\xi
-\int_{I_2}
\frac{ D(\xi+\mu)  }{|\xi+aH_c(T)+bH_c(T)^2|} \, d\xi \right\}.
\end{eqnarray*}
Here,
\begin{eqnarray*}
I_1
&=&
\left[-\hslash\omega_D-(aH_c(T)+bH_c(T)^2)-\mu_BH_c(T), \;
-\hslash\omega_D-(aH_c(T)+bH_c(T)^2)+\mu_BH_c(T) \right], \\
I_2
&=&
\left[\hslash\omega_D-(aH_c(T)+bH_c(T)^2)-\mu_BH_c(T), \;
\hslash\omega_D-(aH_c(T)+bH_c(T)^2)+\mu_BH_c(T) \right].
\end{eqnarray*}

If $D(\xi+\mu)$ is a constant and does not depend on the variable $\xi$, then $\Delta S=0$. But $D(\xi+\mu)$ is a monotone increasing function, and so
\[
\int_{I_1} 
\frac{ D(\xi+\mu)  }{|\xi+aH_c(T)+bH_c(T)^2|} \, d\xi
-\int_{I_2}
\frac{ D(\xi+\mu)  }{|\xi+aH_c(T)+bH_c(T)^2|} \, d\xi<0.
\]
Moreover,
\[
\frac{\partial f}{\, \partial T}(T, \, H_c(T))<0
\]
by Proposition \ref{prp:f} (3). Therefore,
\[
\Delta S<0.
\]
Thus the entropy is discontinuous at $(T,\, H_c(T))$. So we have shown the following. Note that as mentioned in the introduction, we use the unit where the Boltzmann constant $k_B$ is equal to 1 throughout this paper.

\begin{theorem}
The entropy gap $\Delta S$ is negative and is explicitly given by
\begin{eqnarray*}
\Delta S
&=&
-\frac{1}{4} \frac{\partial f}{\, \partial T \,}(T, \, H_c(T))
\left\{ \int_{I_1} 
\frac{ D(\xi+\mu)  }{|\xi+aH_c(T)+bH_c(T)^2|} \, d\xi
-\int_{I_2}
\frac{ D(\xi+\mu)  }{|\xi+aH_c(T)+bH_c(T)^2|} \, d\xi \right\} \\
&<& 0.
\end{eqnarray*}
Therefore, the entropy in the superconducting state is less than that in the normal state, and so the phase transition with external magnetic field is of the first order.
\end{theorem}

\begin{remark}
Let $T \not= \tau_1$. Here, $\tau_1$ is defined by \eqref{eqn:tau-one}. Then $\Delta S<0$ at $(T,\, H_c(T))$ in the $TH$ plane, and $\Delta S=0$ at $(\tau_1,\, 0)$ in the $TH$ plane. Therefore, there is no entropy gap without external magnetic field. So the phase transition without external magnetic field is not of the first order.
\end{remark}

\section{Discussion and conclusions}

Under the simple approximations \eqref{eqn:ap-one}, \eqref{eqn:ap-two} and \eqref{eqn:ap-three}, we construct the Bogoliubov transformation with external magnetic field, and obtain both the mean field BCS Hamiltonian with external magnetic field (in terms of $\gamma_{k\sigma}$ and $\gamma_{k\sigma}^{\dagger}$) and the BCS-Bogoliubov gap equation with external magnetic field.

Supposing that the density of states and the potential are both constants (see \eqref{eqn:constant}), we then apply the implicit function theorem to the BCS-Bogoliubov gap equation with external magnetic field. We show that there is a unique magnetic field (the critical magnetic field) given by a smooth function of the temperature and that there is also a unique nonnegative solution (the gap function) given by a smooth function of both the temperature and the external magnetic field. Note that each of the critical magnetic field and the gap function is given as the implicit function defined by the BCS-Bogoliubov gap equation with external magnetic field.

We obtain the grand potential with external magnetic field that might include the effect similar to that of the FFLO state. We then show that the entropy in the superconducting state is less than that in the normal state, and obtain the explicit expression  for the entropy gap $\Delta S$. In this way, we show that the transition from the normal state to the superconducting state in a type I superconductor is of the first order. Here the density of states is a function and is not a constant. But, as mentioned just above, we suppose that the density of states and the potential are both constants (see \eqref{eqn:constant}) when we apply the implicit function theorem in order to solve the BCS-Bogoliubov gap equation with external magnetic field. 

Therefore, we have to deal with the case where the density of states and the potential are both functions when we try to solve the BCS-Bogoliubov gap equation with external magnetic field. In this case, we need to use operator-theoretical treatment base on fixed-point theorems. We will study this interesting case in an upcoming paper.

\bigskip

\textbf{Data availability statement} \quad  All data supporting the findings of this study are included within the article.

\bigskip

\textbf{Acknowledgments} \quad This work was supported in part by JSPS Grant-in-Aid for Scientific Research (C) KAKENHI Grant Number JP21K03346.

\bigskip

\textbf{Conflict of interest} \quad The author of this work declares that he has no conflicts of interest.


\end{document}